\documentclass[numbers]{article}
\usepackage{arxiv}

\usepackage{hyperref}

\usepackage[numbers]{natbib}

\usepackage{amsmath,amsfonts}
\usepackage{graphicx}
\usepackage{textcomp}
\usepackage{xcolor}
\usepackage{tikz}
\usepackage{pgfplots}
\usetikzlibrary{arrows,calc,fit,patterns,positioning,shapes,pgfplots.groupplots,decorations.pathreplacing}

\usepackage{enumitem}
\usepackage{booktabs}
\usepackage{algorithm}
\usepackage{listings}
\usepackage[noend]{algpseudocode}
\algnewcommand{\LineComment}[1]{\State \(\triangleright\) #1}

\newcommand{\lstbg}[3][0pt]{{\fboxsep#1\colorbox{#2}{\strut #3}}}
\lstdefinelanguage{diff}{
  basicstyle=\ttfamily\small,
  morecomment=[f][\lstbg{red!20}]-,
  morecomment=[f][\lstbg{green!20}]+,
  morecomment=[f][\lstbg{purple!20}]!,
  morecomment=[f][\textit]{***},
  morecomment=[f][\textit]{---},
}

\usepackage{siunitx}
\newcommand{\numnoround}[1]{\num[round-precision=0]{#1}}

\usepackage{todonotes}

\title{MAGPIE: Machine Automated General Performance Improvement via Evolution of Software}
\renewcommand{\undertitle}{}

\author{%
  \href{https://orcid.org/0000-0003-0485-5279}{\includegraphics[scale=0.06]{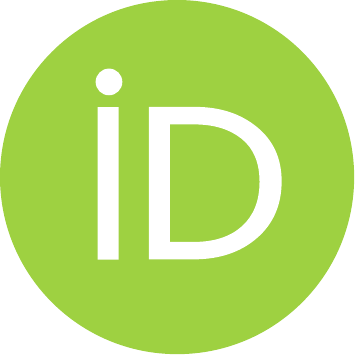}\hspace{1mm}%
  Aymeric~Blot}\\
  Department of Computer Science\\
  University College London\\
  London WC1E 6BT, U.K.\\
  \href{mailto:a.blot@cs.ucl.ac.uk}{a.blot@cs.ucl.ac.uk} \\
  \And
  \href{https://orcid.org/0000-0002-7833-6044}{\includegraphics[scale=0.06]{orcid.pdf}\hspace{1mm}%
  Justyna~Petke} \\
  Department of Computer Science\\
  University College London\\
  London WC1E 6BT, U.K.\\
  \href{mailto:j.petke@ucl.ac.uk}{j.petke@ucl.ac.uk}
}

\date{}

\hypersetup{
  pdftitle={MAGPIE: Machine Automated General Performance Improvement via Evolution of Software},
  pdfsubject={cs.SE},
  pdfauthor={Aymeric~Blot, Justyna~Petke},
  pdfkeywords={parameter tuning, algorithm configuration, genetic improvement, compiler optimisation},
}

\begin{document}

\maketitle

\begin{abstract}
  Performance is one of the most important qualities of software.
  Several techniques have thus been proposed to improve it, such as program transformations, optimisation of software parameters, or compiler flags.
  Many automated software improvement approaches use similar search strategies to explore the space of possible improvements, yet available tooling only focuses on one approach at a time.
  This makes comparisons and exploration of interactions of the various types of improvement impractical.

  We propose MAGPIE, a unified software improvement framework.
  It provides a common edit sequence based representation that isolates the search process from the specific improvement technique, enabling a much simplified synergistic workflow.
  We provide a case study using a basic local search to compare compiler optimisation, algorithm configuration, and genetic improvement.
  We chose running time as our efficiency measure and evaluated our approach on four real-world software, written in C, C\texttt{++}, and Java.

  Our results show that, used independently, all techniques find significant running time improvements: up to 25\% for compiler optimisation, 97\% for algorithm configuration, and 61\% for evolving source code using genetic improvement.
  We also show that up to 10\% further increase in performance can be obtained with partial combinations of the variants found by the different techniques.
  Furthermore, the common representation also enables simultaneous exploration of all techniques, providing a competitive alternative to using each technique individually.

\end{abstract}

\section{Introduction}
\label{section:introduction}

\emph{Software is never done}~\cite{mcquade:2019:dib}.
It needs to be continuously improved. Hidden bugs, misguided assumptions, outdated specifications, code smells, technical debt are all examples of opportunities for software improvement.
Bug-free software is hard to write, and fast, memory and energy-efficient software even more so.
With software present in almost all aspects of our lives its performance has become a key priority.
More than ever, software needs to be faster, more reactive, to be less of a drain on mobile batteries, to use less bandwidth, etc.

There are many ways software performance can be improved.
First of all, programming languages are defined, and shipped with a \emph{default} compiler or interpreter implementing that language.
Alternative compilers or interpreters will often provide different standard-compliant implementations that may lead to performance changes for a given software.
They may also expose optimisation related parameters (e.g., the well known \texttt{-O3} option of C compilers, or garbage collection parameters) that can be specifically tuned to improve execution time~\cite{garciarena:2016:gi-gecco}.
Users can also tune the parameters of the given software itself~\cite{lopez-ibanez:2016:orp}, as software developers will often expose options to allow them to choose different features to be run.
For instance, users might specify a particular output format, a different search algorithm, or a strategy that favours fast but suboptimal results.
Finally, it has been shown that in many cases~\cite{petke:2018:tevc} software can be improved beyond configuring their exposed parameters, by directly modifying their source code.

Automated tooling has been proposed to help developers in automating the software optimisation process.
Examples include the COLE~\cite{hoste:2008:cgo} framework that uses a multi-objective genetic algorithm for GCC compiler optimisation; ParamILS~\cite{hutter:2007:aaai,hutter:2009:jair} that utilises local search to navigate the space of algorithm configurations; and Gin~\cite{brownlee:2019:gecco}, a genetic improvement framework that applies similar search strategies to explore the space of software mutations, with the aim of finding improved software variants.

Albeit originating from completely different fields, compiler optimisation, algorithm configuration, and genetic improvement have all been using similar search strategies, yet have never been compared or combined.
It is unclear whether certain compiler options that improve runtime execution of a program, for instance, would retain the performance gain under different software configuration, that on its own could boost software execution time.
It is also unclear whether such changes are synergistic or not.
Finally, to this point there is no evidence that would support concurrent exploration of the search spaces of compiler or interpreter options, program configurations, and software mutations.

To address this gap, we propose MAGPIE --- \emph{Machine Automated General Performance Improvement through Evolution of software} --- a framework that enables comparison and combination of various software improvement techniques.
At its core, MAGPIE provides a unified software representation that separates the search strategy from the improvement technique.
It allows a natural reuse of search strategies between improvement techniques and their fair comparison, as well as an inherent way of investigating the interactions between improvements found at different granularity levels, e.g., at the level of compiler options or code mutations.

We instantiate MAGPIE with three software improvement techniques: compiler optimisation, algorithm configuration, and genetic improvement.
As the search strategy we use a simple local search as all three techniques have been previously shown to be efficiently solved using dedicated local search strategies.
Focusing on a single search strategy will allow for fair comparison of the magnitude of improvements found by each technique, while future work can explore more complex or different search strategies.

We use our framework to optimise four large real-world software written in C, C\texttt{++}, and Java.
We optimise software performance in terms of running time in three scenarios, while the last scenario exemplifies MAGPIE multi-objective abilities by simultaneously optimising running time and solution quality.
Specifically, we first compare the independent use of compiler optimisation, algorithm configuration, and genetic improvement, before analysing combinations of their recommended modifications, and finally using MAGPIE to simultaneously explore all types of three techniques.

Our results show significant improvements on all four targeted real-world software applications.
In particular, we achieved up to 25\% running time improvement using compiler optimisation, up to 97\% improvement using algorithm configuration, and up to 61\% improvement for evolving source code using genetic improvement.
Automated combination of the patches yielded by applying different techniques led up to 10\% further speedups on specific scenarios.
Finally, whilst joint use of compiler optimisation, algorithm configuration, and genetic improvement was in most cases less efficient than separate training, it was still effective and led to a further 15\% speedup on one scenario.
In summary, our contributions are:

\begin{enumerate}
\item MAGPIE, a unified abstract framework for comparison and combination of diverse software improvement techniques.
\item An instantiation of MAGPIE that combines three ways to improve software performance: compiler options, software parameters, and source code modifications, through an edit sequence representation.
\item The first comparison study of these three software improvement techniques, sharing the same software representation and search strategy.
\item We showed, on four large real-world software, that even the simplest local search can yield significant improvement for all three improvement techniques, comparable with previous work.
  In particular we obtained up to 25\% speedups with compiler or interpreter flag optimisation, up to 61\% for source code modifications evolved using genetic improvement, and up to 97\% via algorithm configuration.
\item We also reported on the first automated exploration of the combination of modifications yielded by different techniques, which showed up to 10\% further improvements.
\item We also conducted a study on the joint search space yielding up to 15\% improvements.
\item Finally, with this paper we provide an implementation of MAGPIE, that provides researchers with quick prototyping and fair comparison of search strategies and improvement techniques, as well as providing practitioners with a single tool to easily apply these improvement techniques. (\url{https://github.com/bloa/magpie})
\end{enumerate}

\section{Related Work}
\label{section:background}

Several techniques have been proposed for improvement of software performance.
These include, but are not limited to, modification of software's source code, (e.g., refactoring~\cite{mens:2004:tse,agnihotri:2020:jips}, genetic improvement~\cite{petke:2018:tevc}), application of optimisations during compilation (e.g., loop transformations~\cite{bacon:1994:acmcs}), use of alternative algorithms or strategies (e.g., algorithm selection~\cite{kotthoff:2014:aimag,kerschke:2019:evco} and configuration~\cite{hamadi:2012,huang:2020:tevc}), making better use of machine specific hardware (e.g., cache management~\cite{gracioli:2015:acmcs}), or even adaptation of the hardware itself on which software is to be run (e.g., with FPGAs~\cite{guo:2019:trts,mittal:2020:nca}).

This leaves software users and developers to choose from a plethora of software improvement options and associated tooling.
We note, however, that several of the aforementioned techniques use similar, if not the same, metaheuristic-based search strategies to navigate the search space of alternative configurations or software modifications.
In this section we elaborate on such techniques.

\subsection{Compiler Optimisation}
\label{subsection:background_compiler}

Human-written code is seldom executed directly.
In most cases, it is interpreted (e.g., with languages such as Python or Ruby) or compiled into machine code or bytecode (e.g., with software such as C, C\texttt{++}, or Java).
The choice of the interpreter or compiler is a first factor that can significantly impact software performance, as the default choice is not guaranteed to be optimal.
For specific applications or environments alternative compilers/interpreters can lead to significant improvements in terms of running time, memory usage, etc.
In particular, compilers such as GCC can offer several hundreds\footnote{\url{https://gcc.gnu.org/onlinedocs/gcc/Optimize-Options.html}} of optimisation options which may interact within each other.
Even if shortcuts to specifically curated \emph{standard} subsets of optimisation options are provided (e.g., \texttt{-O3}), simply enabling more optimisation doesn't always lead to better performance~\cite{georgiou:2018:scopes}.

Research on compiler optimisation follows two major directions: tackling either optimisation selection (choosing which optimisation to apply) or phase-ordering (choosing the order in which to apply optimisations).
Whilst many approaches have been proposed for both, mostly based on machine learning~\cite{ashouri:2019:acmcs}, many metaheuristics have been successfully applied to the arguably simpler problem of optimisation selection.
For instance, multi-objective Strength Pareto Evolutionary Algorithm~\cite{zitzler:2001:tik} was implemented in COLE~\cite{hoste:2008:cgo} and TACT~\cite{plotnikov:2013:iccs}, and a single-objective genetic algorithm was used to navigate the space of compiler flags in ACOVEA~\cite{melnik:2010:grow}.

\subsection{Algorithm Configuration}
\label{subsection:background_tuning}

Beyond performance improvements due to interpreter or compiler configuration, the other major source of potential improvements lies in the configuration of the software itself.
Indeed most programs expose design choices in the form of parameters in order to make them adaptable to particular contexts or applications.
The default values of these parameters are usually set to offer overall good performance across all possible inputs.
The two major types of approaches to improve a software configuration are either to find optimal parameter values before using the software (offline setting) or to update parameter values while using it (online setting)~\cite{eiben:1999:tevc,hamadi:2012}.

In this work, we focus on offline software configuration, usually either called \emph{parameter tuning} or \emph{algorithm configuration}.
While both terms are mostly interchangeable, \emph{tuning} is preferred when all parameters are numerical and continuous (e.g., probability or percentage threshold, number of iterations, population size) leaving \emph{configuration} for when it also involves categorical variables (e.g., strategy selection, Boolean parameters).
\emph{Online} parameter control (and algorithm scheduling~\cite{pageau:2019:emo}) while allowing for further adaptation, comes with the cost of increased search space and complexity.

Most approaches for automated algorithm configuration are metaheuristics.
Well-known automated configurators include for example ParamILS~\cite{hutter:2007:aaai,hutter:2009:jair}, based on local search, SMAC~\cite{hutter:2011:lion}, based on Bayesian modelling, GGA~\cite{ansotegui:2009:cp-2,ansotegui:2015:ijcai-2}, based on a genetic algorithm, and irace~\cite{lopez-ibanez:2016:orp}, based on statistical racing.
Relevant to our research, irace has been directly applied on GCC~\cite{perez-caceres:2017:ea} and SMAC on two prominent JavaScript compilers~\cite{fawcett:2017:corr}, showing that off-the-shelf algorithm configuration can also be used for compiler optimisation.
Another Bayesian approach, BOCA~\cite{chen:2021:icse}, has also been proposed to tackle compiler autotuning and applied on GCC and LLVM.

\subsection{Software Evolution}
\label{subsection:background_gi}

Automated algorithm configuration is usually applied to parameters explicitly exposed by software developers, representing different design choices.
We present below several techniques that operate directly on the software itself, thus allowing for even more software variants to be generated.

Genetic improvement (GI) uses metaheuristics to improve software performance by evolving software itself~\cite{petke:2018:tevc}.
Changes are most commonly applied at the level of source code, inserting or deleting lines, or nodes of an abstract syntax tree (AST).
Early GI work used genetic programming~\cite{koza:1992,langdon:2002,poli:2008} to evolve the target software.
Nowadays the field is expanding, using other types of search approaches, such as local search, recently shown as effective as genetic programming~\cite{blot:2021:tevc}.
GI reuses material already present in the original software, following the plastic surgery hypothesis~\cite{barr:2014:fse}, i.e., that the changes required to obtain improved software can already be found in existing code.
Nonetheless, GI can also use external genetic material, e.g., for automated code transplantation~\cite{barr:2015:issta,petke:2018:tse}.
GI usually mutates lines of codes or statements.
However, it has also been applied to more specific constructs such as Boolean conditionals~\cite{petke:2018:tse}, constants~\cite{langdon:2018:ssbse,krauss:2020:eurogp,langdon:2021:telo}, arithmetic (such as \texttt{<=}/\texttt{<}/\texttt{==}/\texttt{!=}/\texttt{>=}/\texttt{>}), and logical operators (such as \texttt{||}/\texttt{\&\&})~\cite{haraldsson:2017:gi-gecco:1}.

Basios et al. proposed \emph{Darwinian evolution}~\cite{basios:2017:ssbse,basios:2018:fse} which evolves software's data structures.
Their study on Java software identified optimal implementations of List containers for specific variables (e.g., ArrayList, LinkedList).
Wu et al. proposed \emph{deep parameter optimisation} (DPO)~\cite{wu:2015:gecco,bruce:2016:ssbse,bokhari:2017:gi-gecco,white:2017:evoapp} as an alternative solution to overcome the hurdle that software parameters have to be explicitly exposed by software designers.
DPO uses mutation testing as the underlying tool to discover potential unexposed decision choices, before applying multi-objective search for improved parameter values. 
Binkley et al. proposed \emph{observation-based program slicing}~\cite{binkley:2015:scam} which minimises the size of software given a set of criteria to fulfil.
Other approaches include 
\emph{program transformation}~\cite{fatiregun:2003:gecco} and \emph{code refactoring}~\cite{mens:2004:tse} which apply more restrictive template-based changes. 
Another notable example is \emph{loop perforation} for which evolutionary strategies have also been tried~\cite{bankovic:2015:hais}.

\section{The MAGPIE Framework}
\label{section:contribution}

In this section we introduce our conceptual framework MAGPIE --- \emph{Machine Automated General Performance Improvement via Evolution of software} --- and describe how it can be applied to run compiler optimisation, algorithm configuration, and genetic improvement.
To the best of our knowledge we are the first to combine in a unified framework all three types of approaches.

\subsection{MAGPIE Overview}
\label{subsection:overview}

In MAGPIE we propose to represent all changes, from source code changes to configuration changes, as minimal \emph{edits} to be applied to the software, with the final solution representation being the sequence of edits needed to be applied to obtain the desired software variant.
While more complex representations have been proposed~\cite{oliveira:2018:ese}, edit sequences are more easily encoded, constructed, and mutated.
They also make MAGPIE scalable to real-world software.
In addition, Weimer et al. showed that shorter patches are easier to understand and thus more likely accepted by reviewers~\cite{weimer:2006:gpce}.

\begin{figure}[tb]
  \centering
  \begin{tikzpicture}[node distance=3em and 4em,
      block/.style={minimum width=5.5em, minimum height=2em,font=\small},
      rblock/.style={block, rectangle, draw=black, rounded corners},
      arrow/.style={-latex, thick, shorten >=0.5em, shorten <=0.5em},
      every text node part/.style={align=center}]
    \node (N3) [rblock] {\bf MAGPIE};
    \node (N1) [block, left=of N3] {Software\\Benchmark};
    \node (N4) [rblock,below=of N3] {Training set};
    \node (N5) [block,right=of N3] {Best variant};
    \draw [arrow] ([yshift=0.5em] N1.east) to ([yshift=0.5em] N3.west);
    \draw [arrow] ([yshift=-0.5em] N1.east) to ([yshift=-0.5em] N3.west);
    \draw [arrow, bend left=45] (N3) to node [right,xshift=0.2em,font=\footnotesize] {software + \\edit sequence} (N4);
    \draw [arrow, bend left=45] (N4) to node [left,xshift=-0.2em,font=\footnotesize] {fitness} (N3);
    \draw [arrow] (N3) -- (N5);
  \end{tikzpicture}
  \caption{MAGPIE general workflow}
  \label{tikz:magpie}
\end{figure}
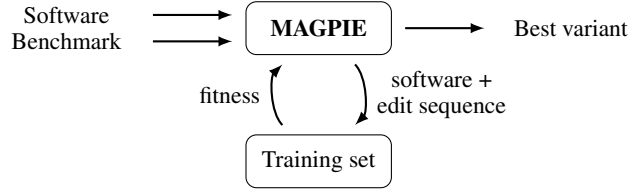

Starting with an initial software and a benchmark on which to improve its performance, MAGPIE will iteratively create edits and construct new software variants to be evaluated, and serve as a basis for subsequent variants.
A given search process then navigates the space of available software edits and ultimately returns the sequence of edits describing how to obtain the best variant found.
A basic workflow of MAGPIE is provided in \autoref{tikz:magpie}.

As frequently used in previous work (see~\autoref{section:background}), we recommend the use of metaheuristics, based on local search~\cite{hoos:2004} or genetic programming~\cite{koza:1992,langdon:2002}.

\subsection{General-Purpose Software Edits}
\label{subsection:magpie_edits}

In MAGPIE each edit is represented as a triplet of elements:
\begin{itemize}[nosep]
\item \emph{edit type:} the type of operation being applied,
\item \emph{location:} where the operation will be applied to, and
\item \emph{ingredient:} if necessary, a new value or code fragment.
\end{itemize}

This type of solution representation has previously been shown to be effective to represent software variants in the context of genetic improvement.
However, it is abstract enough to easily encode other types of software improvement techniques.
Edits can be applied at different granularity levels, summarised as follows.
\begin{itemize}[nosep]
\item \emph{Compiler/interpreter} level, targeting parameters outside the scope of the software source code, e.g., compiler or interpreter options.
\item \emph{Software parameter} level, targeting all accessible developer-exposed software parameters.
\item \emph{Source code} level, targeting lines, statements, e.g., their insertion/deletion/replacement; and mutations of numerical literals, operators, data structures, and others.
\end{itemize}

We describe below how various software improvement techniques can be represented. 
To consider more diverse software in our experiments we focus on edit types that can be applied on most software.
For example, we consider modifying numerical constants but not specific data structures, as those depend on the software programming language.
Nevertheless, we plan to enrich MAGPIE with such specialised program transformations in future work.

\subsection{Compiler/Interpreter Optimisation}
\label{subsection:pyggi_compiler}

Compiler optimisation, as in the selection of the best optimisations to be applied during compilation, is straightforward using algorithm configuration.
Therefore, we use the edit representation described in~\autoref{subsection:pyggi_config}.
Configuring other environmental options, such as the parameters of the language interpreter, are also covered by general algorithm configuration.

In practice edits targeting the configuration of the compiler are used when the software is compiled without interfering with the edits targeting the interpreter or the software itself.

\subsection{Algorithm Configuration}
\label{subsection:pyggi_config}

A typical algorithm configuration solution representation is an associative array that maps every parameter to its corresponding selected parameter value.
Instead, we propose to encode the changes to the default configuration rather than evolving complete copies.
Therefore, each \emph{replacement edit} for algorithm configuration specifies the parameter it targets (\emph{location}) together with its new value (\emph{ingredient}).

Parameter values can be drawn uniformly at random from the range of possible values, but also following other statistical distributions (e.g., geometric) or combinations of any distributions (e.g., specific values such as $0$ or $-1$ being treated separately).
Special cases such as conditional parameters and forbidden combinations are handled during fitness evaluation ensuring the validity of the resulting configuration.

The number of possible edits is the sum of the number of possible parameter values for all parameters.
This becomes unreasonably large as soon as a parameter can assume floating point numerical values or admit large bounds (e.g., an 32-bit unsigned integer admits $2^{32}$ possible values).
However, at any point edits can still be enumerated by subsampling for each such parameter a set number of possible values.

\subsection{Genetic Improvement}
\label{subsection:pyggi_gi}

Genetic improvement (GI) typically targets source code, mutating either lines of code or statements~\cite{petke:2018:tevc}.
For the latter the abstract syntax tree (AST) representation is used.
There are three standard GI \emph{edit types}: deletion, replacement, and insertion, each specifying code fragments as both \emph{location} and \emph{ingredient}.
Deletions result in a linear number of possible edits, while numbers of possible replacements and insertions grow quadratically. 

\subsection{Numerical Constants}
\label{subsection:pyggi_data}

The space of possible numerical values is too large to simply sample random replacements.
Instead, we propose two \emph{edit types}, using absolute and relative values.
As for configuration edits, edits modifying numerical constants specify the \emph{location} of the specific constant they target and it modified expression (\emph{ingredient}).

We single out the values $0$, $1$, and $-1$, following Wu et al.~\cite{wu:2015:gecco}, as they are used in mutation testing.
As for relative values, previous work showed that $\pm$1 increments in software numerical constants helped in the context of bug fixing~\cite{haraldsson:2017:gi-gecco:1}.
To enable changes of larger magnitude while still keep replacements with scale related to the original value, we propose the following simple operations: \texttt{(\textbullet+1)}, \texttt{(\textbullet-1)}, \texttt{(\textbullet*2)}, \texttt{(\textbullet/2)}, \texttt{(\textbullet{}*3/2)}, or \texttt{(\textbullet{}*2/3)}.
Operations are applied with correct parenthesising, ensuring that successive edits can stack on top of each other.
We enforce the constraint that a fixed number of new expressions can be applied to every numerical constant.
This way the number of possible edits is linear in the number of constants.

\section{Research Questions}
\label{section:researchquestions}

Our main research goal is to show the benefits of a unified framework for software improvement technologies.
With that in mind, we propose MAGPIE and ask the following RQs:


\textbf{RQ1: How effective is MAGPIE at optimising compiler or interpreter configurations?}
Firstly, we want to measure the effect of different settings of the compiler (or interpreter, depending on the language). This includes, e.g., for C code, the choice of the compiler and automated selection of different parameter settings.

\textbf{RQ2: How effective is MAGPIE at tuning parameters explicitly exposed to the user?}
We want to investigate what improvements can be achieved using automated algorithm configuration.

\textbf{RQ3: How effective is MAGPIE at finding efficiency improvements using genetic improvement?}
Similarly, we want to know what type of improvements could be found manipulating the program's source code, both its statements and numerical values.

\textbf{RQ4: What is the impact of combining efficiency improving changes from different granularity levels?}
The search spaces of compiler optimisation, algorithm configuration, and genetic improvement are diverse in size, structure, and content.
If changes produced by each technique are good in isolation, can performance be further improved by combining them together?

\textbf{RQ5: How effective is MAGPIE at simultaneously exploring the joint search space of software edits?}
Since the edit sequence solution representation is shared within MAGPIE, despite their differences, can compiler optimisation, algorithm configuration, and genetic improvement be applied at the same time to evolve software?
We also measure the impact of such improvements found.

\section{Empirical Study}
\label{section:protocol}

In order to answer our research questions we conducted an empirical study on four large real-world software.
In this section we present details of our experimental protocol, the MAGPIE implementation, and the benchmarks used.

\subsection{Experimental Protocol}
\label{subsection:protocol}

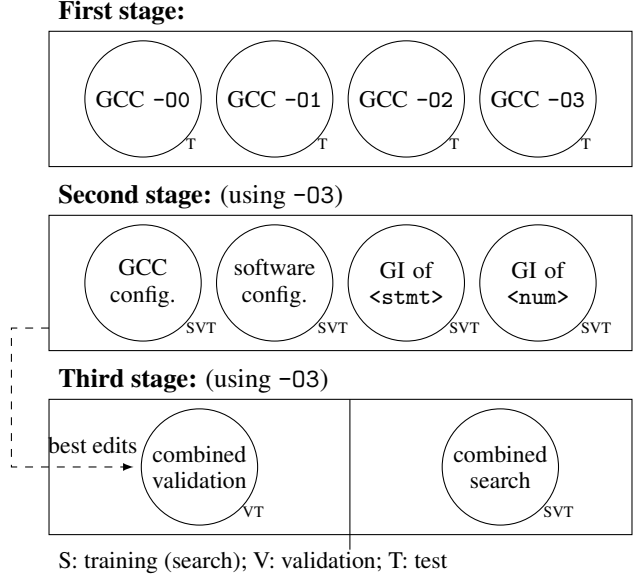
\begin{figure}[tb]
  \centering
  \begin{tikzpicture}[
      shortnode/.style={text width=3.65em,align=center,font=\small},
      mininode/.style={anchor=north west,font=\tiny},
      b0/.style={anchor=mid,text height=1.5ex,text depth=0.25ex},
      b1/.style={b0,draw,ellipse},
      b2/.style={b0,draw,rectangle,minimum height=1.5em},
    ]
    \node (bl0) {};
    \node (bl1) [below=2.2 of bl0] {};
    \node (bl2) [below=2.2 of bl1] {};

    \node [anchor=west] at ($(bl0)+(0,0.25)$) {\textbf{First stage:}};
    \node [anchor=west] at ($(bl1)+(0,0.25)$) {\textbf{Second stage:} (using \texttt{-O3})};
    \node [anchor=west] at ($(bl2)+(0,0.25)$) {\textbf{Third stage:} (using \texttt{-O3})};

    \node at ($(bl0)+(1.25,-0.9)$) [shortnode] {GCC \texttt{-O0}};
    \node at ($(bl0)+(1.25,-0.9)+(0.45,-0.4)$) [mininode] {T};
    \node at ($(bl0)+(3,-0.9)$) [shortnode] {GCC \texttt{-O1}};
    \node at ($(bl0)+(3,-0.9)+(0.45,-0.4)$) [mininode] {T};
    \node at ($(bl0)+(4.75,-0.9)$) [shortnode] {GCC \texttt{-O2}};
    \node at ($(bl0)+(4.75,-0.9)+(0.45,-0.4)$) [mininode] {T};
    \node at ($(bl0)+(6.5,-0.9)$) [shortnode] {GCC \texttt{-O3}};
    \node at ($(bl0)+(6.5,-0.9)+(0.45,-0.4)$) [mininode] {T};
    \draw ($(bl0)+(1.25,-0.9)$) circle (2.2em);
    \draw ($(bl0)+(3,-0.9)$) circle (2.2em);
    \draw ($(bl0)+(4.75,-0.9)$) circle (2.2em);
    \draw ($(bl0)+(6.5,-0.9)$) circle (2.2em);

    \draw (bl0) rectangle ++(7.75,-1.8);
    \draw (bl1) rectangle ++(7.75,-1.8);
    \draw (bl2) rectangle ++(7.75,-1.8);
    \draw [-latex,dashed] ($(bl1)+(0,-1.5)$) -- ++(-0.5,0) |- ($(bl2)+(1.125,-0.9)$) node [anchor=south east,yshift=0.15em,xshift=0.5em,font=\footnotesize] {best edits};

    \node at ($(bl1)+(1.25,-0.9)$) [shortnode] {GCC config.};
    \node at ($(bl1)+(1.25,-0.9)+(0.45,-0.4)$) [mininode] {SVT};
    \draw ($(bl1)+(1.25,-0.9)$) circle (2.2em);
    \node at ($(bl1)+(3,-0.9)$) [shortnode] {software config.};
    \node at ($(bl1)+(3,-0.9)+(0.45,-0.4)$) [mininode] {SVT};
    \draw ($(bl1)+(3,-0.9)$) circle (2.2em);
    \node at ($(bl1)+(4.75,-0.9)$) [shortnode] {GI of \texttt{<stmt>}};
    \node at ($(bl1)+(4.75,-0.9)+(0.45,-0.4)$) [mininode] {SVT};
    \draw ($(bl1)+(4.75,-0.9)$) circle (2.2em);
    \node at ($(bl1)+(6.5,-0.9)$) [shortnode] {GI of \texttt{<num>}};
    \node at ($(bl1)+(6.5,-0.9)+(0.45,-0.4)$) [mininode] {SVT};
    \draw ($(bl1)+(6.5,-0.9)$) circle (2.2em);

    \node at ($(bl2)+(2,-0.9)$) [shortnode] {combined validation};
    \node at ($(bl2)+(2,-0.9)+(0.45,-0.4)$) [mininode] {VT};
    \node at ($(bl2)+(6,-0.9)$) [shortnode] {combined search};
    \node at ($(bl2)+(6,-0.9)+(0.45,-0.4)$) [mininode] {SVT};
    \draw ($(bl2)+(2,-0.9)$) circle (2.2em);
    \draw ($(bl2)+(6,-0.9)$) circle (2.2em);
    \draw ($(bl2)+(4,-2)$) -- ++(0,2);

    \node at ($(bl2)+(0,-1.9)$) [anchor=north west,font=\footnotesize] {S: training (search); V: validation; T: test};

  \end{tikzpicture}
  \caption{Experimental protocol (for C/C\texttt{++} and GCC).
    In the first stage compiler optimisation levels are manually evaluated.
    In the second stage four independent runs are conducted.
    Each only use different types of edits, corresponding to four approaches: compiler optimisation, algorithm configuration, mutation of statements, mutation of numerical literals.
    In the third stage we evaluate the combination of the best edits found in the second stage.
    We also evolve software using every available edit type, conceptually combining all four approaches.}
  \label{figure:protocol}
\end{figure}

The general workflow of the experiments is illustrated in \autoref{figure:protocol}.
For simplicity, we illustrate our procedure on an example using a C/C\texttt{++} software and the GCC compiler.

First of all, we manually investigate a limited number of specific external settings.
For C/C\texttt{++} software we use the compiler optimisation parameters \texttt{-O0} to \texttt{-O3}, whereas for Java software we simply use each JVM with default settings.

Then, we investigate how much performance of a piece of software can change when the parameters of the compiler or the virtual machine are optimised.
Next, we investigate the impact of configuring the software itself.
In particular, we run independent evolution of software's parameters, statements and numerical values.
For C/C\texttt{++} evolution starts from \texttt{-O3} to minimise the length of the experiments: we chose to not use the previously evolved compiler configuration, even if better performance was found, to avoid dependencies between experimental steps.

Finally, in the last step we evaluate the effect of combining edits from all the different approaches.
We investigate the following two strategies: 1) we simply combine modifications we obtained during training in the second stage 2) we re-run MAGPIE on the joint edit search space, at each step of search picking an edit type at random, regardless of whether it's applicable to compiler options or numerical constants or other.

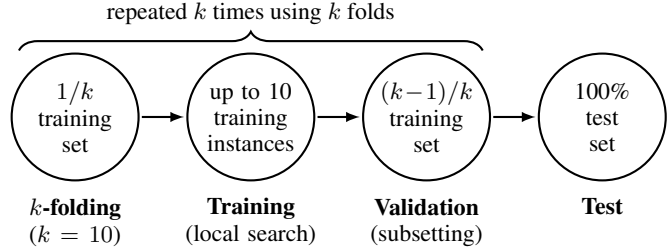
\begin{figure}[tb]
  \centering
  \begin{tikzpicture}[
      shortnode/.style={text width=3.5em,align=center,font=\small},
      longnode/.style={text width=5em,align=center,font=\small},
      thick,
    ]
    \node (circle1) [shortnode] {$1/k$ training set};
    \node (circle2) [shortnode,right=2.4em of circle1] {up to 10 training instances};
    \node (circle3) [shortnode,right=2.4em of circle2] {$(k-1)/k$ training set};
    \node (circle4) [shortnode,right=2.4em of circle3] {100\%\\ test\\ set};
    \draw (circle1) circle (2.4em);
    \draw (circle2) circle (2.4em);
    \draw (circle3) circle (2.4em);
    \draw (circle4) circle (2.4em);

    \node [longnode,below=1.0em of circle1] {\textbf{$k$-folding} ($k=10$)};
    \node [longnode,below=1.0em of circle2] {\textbf{Training} (local search)};
    \node [longnode,below=1.0em of circle3] {\textbf{Validation} (subsetting)};
    \node [longnode,below=1.0em of circle4] {\textbf{Test}};

    \draw [-latex] ($(circle1.east)+(0.4em,0)$) -- ($(circle2.west)+(-0.4em,0)$);
    \draw [-latex] ($(circle2.east)+(0.4em,0)$) -- ($(circle3.west)+(-0.4em,0)$);
    \draw [-latex] ($(circle3.east)+(0.4em,0)$) -- ($(circle4.west)+(-0.4em,0)$);

    \draw[decorate,decoration={brace,amplitude=0.25cm}]  ($(circle1.north west)+(0em,0.8em)$) -- ($(circle3.north east)+(0em,0.8em)$) node [midway,yshift=1.4em] {\small repeated $k$ times using $k$ folds};
  \end{tikzpicture}
  \caption{Cross-validation with $k$-folding and holdout.}
  \label{figure:validation}
\end{figure}

We use actual software execution to evaluate the fitness of software variants (Figure 1).
To make sure performance improvements generalise on unseen data, available instances are divided into two disjoint subsets: a \emph{training set}, to compare variants to one another; and a \emph{test set}, subsequently used to confirm the quality of the final best software variant found.
In addition, to make sure results are not tainted by the choice of the training set, we use $k$-fold cross-validation in addition to simple holdout, as illustrated in \autoref{figure:validation}.
Training instances are then separated again into $k$ subsets so that training is repeated $k$ times.
For each trial of our search algorithm only 10 training instances are actually used due to the very high evaluation cost.
At the end of the training step, every instance from the other $k-1$ subsets are used to validate the final software variant.
Finally, after the $k$ training and validation repetitions, the single best program variant is evaluated on the test set.
That scheme is used for every step of the experimental protocol (\autoref{figure:protocol}).

\subsection{MAGPIE Implementation}
\label{subsection:pyggi}

To allow for language-independent experiments, we instrumentalise our implementation with a universal XML representation of software on which to apply our edit sequence representation.
We chose to use srcML~\cite{collard:2016:icsme} to convert source code to an XML representation.
srcML breaks down a source code file into its abstract syntax tree (AST) resulting in an easy to modify structure.
It provides provides support for C, C\texttt{++}, C\texttt{\#}, as well as Java.
This approach was previously successfully used in the PyGGI~2.0~\cite{an:2019:fse} GI tool. 

In addition, to represent compiler or software parameters configurations, we used dictionary-like linear tree structures for a natural and seamless integration.
We implemented the following edits:
\begin{description}
\item[\texttt{StmtDelete}] that deletes a statement node from the AST;
\item[\texttt{StmtReplace}] that replaces a statement node in the AST by another from the original AST;
\item[\texttt{StmtInsert}] that inserts (before or after another statement node) a statement node from the original AST;
\item[\texttt{ConstantSet}, \texttt{ConstantUpdate}:] that modifies a numerical constant according to the ingredient content;
\item[\texttt{ParamSet}] that assigns to a (either compiler or software) parameter the value passed as ingredient.
\end{description}

\begin{algorithm}[tb!]
  \begin{algorithmic}[]
    \Procedure{$\mathit{LS}$}{\,}
      \State $\mathit{best}\gets$ empty mutant
      \Repeat
        \LineComment{Append or remove an edit at random}
        \State $\mathit{mutant}\gets \mathit{mutate}(\mathit{best})$
        \LineComment{Accept if better}
        \If{$\mathit{fitness(mutant)} \leq \mathit{fitness(best)}$}
          \State $\mathit{best}\gets \mathit{mutant}$
        \EndIf
      \Until{training budget exhausted}
      \State \textbf{return} $\mathit{best}$
    \EndProcedure
  \end{algorithmic}
  \caption{First improvement local search}
  \label{algo:first_improvement}
\end{algorithm}

Following recent results~\cite{blot:2021:tevc}, we use a local search procedure (see~\autoref{algo:first_improvement}).
The initial solution is an empty sequence, corresponding to the unmodified software.
Then, edits, drawn uniformly at random, are increasingly appended to the current solution as long as performance is not negatively impacted.
Alternatively, mutation can also remove any edit previously inserted in the edit sequence.

As for the validation step, the edit sequence needs to be reduced to a minimal form, in order to both avoid overfitting on the training set and to provide a simpler final patch.
With the assumption that edits are mostly independent, especially when combining edits obtained during separate runs (\autoref{figure:protocol}, third step: combined validation), all edits are first independently evaluated and ranked with respect to runtime efficiency.
Then, they are reintroduced one by one and appended to the new edit sequence and discarded unless they contribute to further performance improvement~\cite{blot:2020:eurogp}.
However, due to edit interactions, sometimes such constructed sequence is unable to match the performance of the naive concatenation of all edits.
In this case, we start from the complete sequence and successively remove each edit from the sequence until no single edit can be removed without performance loss~\cite{langdon:2015:tevc}.

Training trials were run in parallel using a budget of 1000 mutant evaluations, except for the final step that uses a budget of 4000 evaluations to accommodate for the much larger joint search space.
Cross-validation uses the standard $k=10$ repetitions.

\subsection{Compiler and VM options}
\label{subsection:setup}

We chose to compare GCC, the defacto-standard linux C compiler, to Clang, a more recent compiler.
Whilst by far most of the literature concentrate on GCC, which exposes a very large number of options, recent work has also been focusing on LLVM/Clang~\cite{colucci:2020:corr,chen:2021:icse,georgiou:2022:cj,li:2022:gi-gecco}.
Both compilers are known to produce semantically similar yet different machine code, potentially leading to differences in performance.
Whilst both compilers expose different optimisation options, they both provide the \texttt{-O0} to \texttt{-O3} optimisation shortcuts.

Whilst compiler optimisation typically only targets compiled languages such as C, we also compare the OpenJDK and GraalVM Java virtual machines (JVM) with the expectation that different execution heuristics will also lead to different performance.
While optimisation options are significantly less numerous than for C compilers, JVMs provide many extra options\footnote{\url{https://docs.oracle.com/en/java/javase/14/docs/specs/man/java.html}} that are subject to change without notice and therefore hard to tune manually.
The choice of considering Java software is also motivated by a recent report on the JVM ecosystem highlighting that 36\% of surveyed developers switched from the Oracle JDK to an alternative OpenJDK alternative in 2019.\footnote{\url{https://snyk.io/blog/jvm-ecosystem-report-2020/}}
Details regarding the two C/C\texttt{++} compilers and two JVMs as succinctly presented below.

\begin{description}
\item[GCC 9.3.1]
GCC is the standard GNU and Linux compiler.
In addition to the usual \texttt{-O0} to \texttt{-O3} options we consider 213 other Boolean and categorical optimisation parameters.

\item[Clang 12.0.0]
Clang is a compiler based on the LLVM toolchain, designed to provide a drop-in replacement to GCC.
While Clang has made an explicit design decision not to expose the optimisation pipeline details to consumers, there are 63 transform passes that can be enabled in order to optimise the final executable.

\item[OpenJDK 12.0.2 (Oracle)]
  We use a recent (2019) and widespread Java implementation as a baseline.
  We use 96 parameters shared between both Java implementations, including VM runtime, JIT compiler, and garbage collection options.

\item[OpenJDK 11.0.9 (GraalVM 20.3.0)]
  While based on a slightly older Java version (2018), GraalVM is an alternative Java VM and JDK that advertise high-performance running time, fast startup and low memory footprint.
  We use the same parameters as the Oracle-provided Java implementation.
\end{description}

\subsection{Software For Improvement}
\label{subsection:scenarios}

Our experiments require software on which all three types of software improvement techniques can be applied, more specifically C/C\texttt{++} and Java software with relevant exposed software parameters with reasonable compilation time.
We also want to choose software for which we know there are improvements to be found.
To that end, we use the AClib~\cite{hutter:2014:lion}, a benchmark library for software algorithm configuration.
We chose to focus on the 3 largest categories: SAT solving, machine learning and planning software.

The AClib library is dominated by different variants of SAT solvers. 
Two of these have previously been used as case studies in literature on genetic improvement~\cite{blot:2021:tevc,petke:2018:tse}.
In order to ensure diversity in the selected software we also chose two software from different AClib categories: planning and machine learning.
Details regarding the selected software is presented below:

\begin{description}
\item[MiniSAT] (C\texttt{++}) is a well known SAT solver.
  We use the variant \texttt{minisat\_HACK\_999ED\_CSSC-cssc14} as provided by AClib.
  It exposes 25 mostly numerical parameters, and during search we evolve the \texttt{core/Solver.cc} file.
  We use the CircuitFuzz dataset, with 247 training and 277 test instances.

\item[LPG] (C) is a local-search based planning solver.
  We use the latest version of LPG (1.2) as provided by AClib.
  It exposes 87 numerical and categorical parameters.
  Evolution focuses on the \texttt{LocalSearch\/H\_relaxed.c} file.
  We use the Blocksworld dataset, with 49 training and 25 test instances.

\item[Sat4J] (Java) is another SAT Solver, but written in Java.
  We use the latest version of Sat4J (2.3.1 snapshot \texttt{016981c0}) with the 10 categorical and numerical parameters provided by AClib.
  The functions from \texttt{minisat/core/Solver.java} are evolved during the experiments.
  We use the same CircuitFuzz instances as in the MiniSAT scenario.

\item[Weka] (Java) is a popular data mining and machine learning Java software.
  We use the latest version of Weka (3.8 snapshot \texttt{d5ade95}) and focus on the random forest classifier.
  Its 10 parameters and implementations of both the \texttt{RandomForest} and \texttt{RandomTree} classes are optimised.
  We use the Dexter dataset used by Auto-Weka~\cite{thornton:2013:kdd}, with 420 training and 180 test instances.
  Since Weka already performs cross-validation internally, training in our experimental protocol uses every training instance.
\end{description}

Our objective is to improve efficiency in terms of running time with MAGPIE.
Henceforth a technique is considered effective if it produces a software variant that is faster than original software.
However, in practice we chose to measure the number of CPU instructions, as we found it to be less flaky than raw elapsed time (see~\autoref{subsection:rq1}).
Measurements are obtained using the \texttt{perf} Linux kernel performance analysis tool.

Weka computes machine learning models with intrinsic quality in terms of classification rate.
To ensure our Weka variants compute both efficient and fast models, we minimise both misclassification rate and running time in lexicographical order, thus including a bi-objective scenario in our experiments.

\section{Results and Analysis}
\label{section:results}

In this section we present the results of our empirical study and provide answers to our research questions.
All experiments were conducted on a computational cluster running CentOS 7.8.2003 and the Linux 3.10.0 kernel.

\subsection{RQ1: Compiler and JVM Optimisation}
\label{subsection:rq1}

\begin{table}[t!]
  \caption{Average and best fitness on test instances with different compiler and JVM settings. [in CPU instructions]}
  \centering
  \begin{tabular}{l c@{\,}l c@{\,}l}
    \toprule
     & \multicolumn{2}{c}{GCC} & \multicolumn{2}{c}{Clang}\\
    \midrule
    \multicolumn{2}{l}{\textbf{MiniSAT}}\\
    \texttt{-O0} & $\num{84336965193275}$ & $\pm\numnoround{697801.4594110722}$ & $\num{86059099746339}$ & $\pm\numnoround{21015620.934655555}$\\
    \texttt{-O1} & $\num{8978560170820.5}$ & $\pm\numnoround{2585840.665955153}$ & $\num{41560392680353.2}$ & $\pm\numnoround{802475.0510069595}$\\
    \texttt{-O2} & $\num{8586747815216.2}$ & $\pm\numnoround{1799851.3217491542}$ & \bf $\num{8429515833776}$ & $\pm\numnoround{1734298.9709438477}$\\
    \texttt{-O3} & $\num{8560025063208.7}$ & $\pm\numnoround{2070231.2096661543}$ & $\num{8681798790364.2}$ & $\pm\numnoround{2162703.6357165333}$\\
    \textsc{magpie} & \bf $\num{8017393420071.6}$ & $\pm\numnoround{89705099710.88747}$ & $\num{8505166050813.5}$ & $\pm\numnoround{121808288901.27841}$\\
    \cmidrule(lr){1-5}
    best & $\num{7806500120218}$ & (-90.74\% \texttt{-O0}) & $\num{8429512315834}$ & (-90.20\% \texttt{-O0})\\
    && (-8.80\% \texttt{-O3}) && (-2.91\% \texttt{-O3})\\
    \cmidrule(lr){1-5}
    \multicolumn{2}{l}{\textbf{LPG}}\\ 
    \texttt{-O0} & $\num{34829694890915}$ & $\pm\numnoround{10176942.694457276}$ & $\num{27869964538567.9}$ & $\pm\numnoround{18510413.66960084}$\\
    \texttt{-O1} & $\num{17654552804765.1}$ & $\pm\numnoround{276676.95186442416}$ & $\num{13721399049225.3}$ & $\pm\numnoround{4872597.115902521}$\\
    \texttt{-O2} & $\num{13647669591311.8}$ & $\pm\numnoround{3248847.373538492}$ & \bf $\num{12689249488395.6}$ & $\pm\numnoround{3186670.1830262747}$\\
    \texttt{-O3} & $\num{13884363297634.4}$ & $\pm\numnoround{3851512.33271527}$ & \bf $\num{12678457819901.7}$ & $\pm\numnoround{4340388.356515886}$\\
    \textsc{magpie} & \bf $\num{12526988676227.111}$ & $\pm\numnoround{355175098045.47925}$ & \bf $\num{12652233817327.4}$ & $\pm\numnoround{9213346196.716425}$\\
    \cmidrule(lr){1-5}
    best & $\num{12202547786121}$ & (-64.97\% \texttt{-O0}) & $\num{12649319637618}$ & (-54.61\% \texttt{-O0})\\
    && (-12.11\% \texttt{-O3}) && (-0.23\% \texttt{-O3})\\
    \bottomrule
    \toprule
     & \multicolumn{2}{c}{Oracle OpenJDK} & \multicolumn{2}{c}{GraalVM}\\
    \midrule
    \multicolumn{2}{l}{\textbf{Sat4j}}\\
    default & $\num{41658295567448.8}$ & $\pm\numnoround{176759056120.65918}$ & $\num{44268098376402.1}$ & $\pm\numnoround{707162758631.008}$\\
    \textsc{magpie} & \bf $\num{34803722740485.4}$ & \bf $\pm\numnoround{772607250710.0184}$ & \bf $\num{33353871890151.3}$ & $\pm\numnoround{613847618705.6089}$\\
    \cmidrule(lr){1-5}
    best & $\num{33826046917684}$ & (-18.8\% def.) & $\num{33088059927089}$ & (-25.26\% def.)\\
    \cmidrule(lr){1-5}
    \multicolumn{2}{l}{\textbf{Weka}}\\
    default & $\num{651424899990.1}$ & $\pm\numnoround{1709368047.9481494}$ & $\num{654652237958.6}$ & $\pm\numnoround{2675557195.291855}$\\
    \textsc{magpie} & \bf $\num{621746374934.4}$ & $\pm\numnoround{2489680985.102262}$ & \bf $\num{627916495260}$ & $\pm\numnoround{678833208.8901786}$\\
    \cmidrule(lr){1-5}
    best & $\num{619757198178}$ & (-4.86\% def.) & $\num{626856342037}$ & (-4.25\% def.)\\
    \bottomrule
    \label{table:external_factors}
  \end{tabular}
\end{table}

\autoref{table:external_factors} shows, for all four software, the average number of CPU instructions recorded over ten runs (with standard deviations) when run on the full set of test instances.
These are recorded for the default software with standard compiler and VM options, as well as MAGPIE-evolved compiler and VM configurations.
Additionally, the ``best'' row reports the single best fitness produced by MAGPIE.

We first note the extremely low coefficients of variation (ratio of standard deviation over mean) for repeated measurements, despite the relatively long CPU times.
As an illustration, for MiniSAT with GCC \texttt{-O0}, the ten trials took in average \SI{12370}{s} with a standard deviation of \SI{361}{s}, meaning a coefficient of variation of $0.03$, five orders of magnitude larger than \num[scientific-notation=true]{8.273969282768484e-07} when measuring CPU instructions.
For that reason, all performance improvements are reported using ratios of CPU instructions rather than ratios of CPU time.
While the most probable explanation is latency when accessing input files during execution, accumulations of kernel interrupts, context switches, and hidden multi-threading may also contribute to noise in time measurements, especially for the JVMs.

Before even running MAGPIE we observe a clear gap between the use of \texttt{-O0} (default parameter value) and \texttt{-O2}/\texttt{-O3} for both GCC and Clang, with around 90\% speedups for MiniSAT and 40\% to 45\% speedups for LPG.
Improvements upon \texttt{-O2} or \texttt{-O3} using MAGPIE are smaller, up to 12\% speedups for GCC and 3\% speedups for Clang, the latter being due to the unfortunate closed nature of the optimisation options of Clang.
The magnitude of these results is in line with previous work using automated algorithm configuration on GCC~\cite{perez-caceres:2017:ea}.
For Sat4j the best performance is achieved by configuring GraalVM, for which a 25\% speedup in CPU instruction was achieved.
On the more recent Oracle OpenJDK a 19\% speedup was still achieved.
For Weka results for both JVMs are similar, with a 5\% speedup obtained for the Oracle OpenJDK.

\begin{table}[t!]
  \caption{Frequent compiler and interpreter individual parameter changes with $\ge1\%$ improvement in fitness.}
  \centering
  \begin{tabular}{l@{~}lllr}
    \toprule
    Scenario && Parameter & \# & Imp.\\
    \midrule
    MiniSAT & GCC & \texttt{-ftree-loop-im} & 7 & -3.6\%\\
            &     & \texttt{-fira-algorithm=CB} & 6 & -1.3\%\\
    \cmidrule(lr){2-5}
            & Clang & \texttt{-O2} & 7 & -3.0\%\\
    \cmidrule(lr){1-5}
    LPG     & GCC & \texttt{-ftree-loop-im} & 8 & -3.2\%\\
            &     & \texttt{-fivopts} & 7 & -4.7\%\\
            &     & \texttt{-fno-tree-fre} & 6 & -1.0\%\\
    \cmidrule(lr){1-5}
    Sat4j   & Oracle & \texttt{-XX:-UseCompressedOops} & 7 & -2.9\%\\
            &        & \texttt{-XX:+AggressiveHeap} & 6 & -10.5\%\\
            &        & \texttt{-batch} & 5 & -1.5\%\\
    \cmidrule(lr){2-5}
            & GraalVM & \texttt{-XX:-UseCompressedOops} & 10 & -8.4\%\\
            &         & \texttt{-XX:-UseTLAB} & 8 & -1.2\%\\
            &         & \texttt{-XX:+AggressiveHeap} & 5 & -10.9\%\\
    \bottomrule
    \label{table:frequent_external_tuning}
  \end{tabular}
\end{table}

Details for frequently changed parameters with noticeable impact in isolation are presented in \autoref{table:frequent_external_tuning}, as found during the validation step.
Results for Weka and LPG+Clang are omitted as no single individual parameter change had a $>1\%$ improvement.
Interestingly for Java, for which JVM configuration is often advised against, we note a frequent change in the garbage collection algorithm for Sat4j, and the disabled use of \emph{compressed ordinary object pointer}s specifically for GraalVM.

\smallskip
\noindent
\centerline{
\fbox{
  \parbox{0.96\linewidth}{
    \textbf{Answer to RQ1:}
    The edit sequence representation is both effective and efficient at optimising the configuration of both C/C\texttt{++} compilers and Java VMs.
    Speedups up to 12\% were achieved for GCC when compared with \texttt{-O3} (91\% speedup when compared with \texttt{-O0}), 19\% for the Oracle OpenJDK, and 25\% for GraalVM.
  }
}
}

\subsection{RQ2: Algorithm Configuration; RQ3: Genetic Improvement}
\label{subsection:rq2}

\begin{figure}[tb!]
  \small
  \centering
  \begin{tikzpicture}
    \begin{axis}[
        scale only axis,
        xmin=-0.5,
        xmax=9.5,
        minor y tick num=1,
        tick scale binop=\times,
        grid=both,
        xtick={0,1,2,3,4,5,6,7,8,9},
        xlabel={Repetition},
        ylabel={CPU instructions},
        width=0.75\linewidth,
        height=17em,
      	legend style={at={(1,1)},anchor=north east}]
      \addplot[color=blue,mark=none] coordinates {
        (-0.5, 8560025063208)
        (9.5, 8560025063208)
      };
      \addplot[color=red,dashed,mark=none] coordinates {
        (-0.5, 8681798790364)
        (9.5, 8681798790364)
      };
      \addplot[color=blue,mark=text,text mark=c] coordinates {
        (0, 7653025508590)
        (1, 8345129794644)
        (2, 9863383559624)
        (3, 9899629500279)
        (4, 8718739893705)
        (5, 9922444491091)
        (6, 8545852199125)
        (7, 9736343179405)
        (8, 8545123048716)
        (9, 10782682248663)
      };
      \addplot[color=red,dashed,mark=text,text mark=c] coordinates {
        (0, 9290018215965)
        (1, 9155378086830)
        (2, 9178247399976)
        (3, 8997549542365)
        (4, 9767907916214)
        (5, 7178866275918)
        (6, 8257931917987)
        (7, 8854243811929)
        (8, 8658908130254)
        (9, 10925129914923)
      };
      \addplot[color=blue,mark=text,text mark=s] coordinates {
        (0, 8492038604484)
        (1, 7280103432626)
        (2, 9452180409760)
        (3, 7493785436594)
        (4, 12181222796877)
        (5, 8517291908168)
        (6, 8969719075239)
        (7, 9035149761301)
        (8, 9392987840131)
        (9, 13227407723583)
      };
      \addplot[color=red,dashed,mark=text,text mark=s] coordinates {
        (0, 8717253388709)
        (1, 10838457997291)
        (2, 11140279130425)
        (3, 13305008337582)
        (4, 10759348783575)
        (5, 8870712832441)
        (6, 9409463529104)
        (7, 9214257253116)
        (8, 9760231372066)
        (9, 10039212357215)
      };
      \addplot[color=blue,mark=text,text mark=n] coordinates {
        (0, 8559587707861)
        (1, 8822651725424)
        (2, 9031664429333)
        (3, 9368642553869)

        (6, 8559198496057)

        (8, 10898373389069)
      };
      \addplot[color=red,dashed,mark=text,text mark=n] coordinates {
        (0, 8652192743276)
        (1, 8880176154667)
        (2, 9151538670893)
        (3, 11443038906379)
        (4, 9452446630190)

        (6, 8651672792294)
        (7, 8758274705269)
      };
      \legend{GCC,Clang}
    \end{axis}
  \end{tikzpicture}
  \caption{Algorithm configuration and source code evolution using MAGPIE on the MiniSAT scenario. (c:~configuration, s:~statements, n:~numerical constants)}
  \label{tikz:test_minisat}
\end{figure}
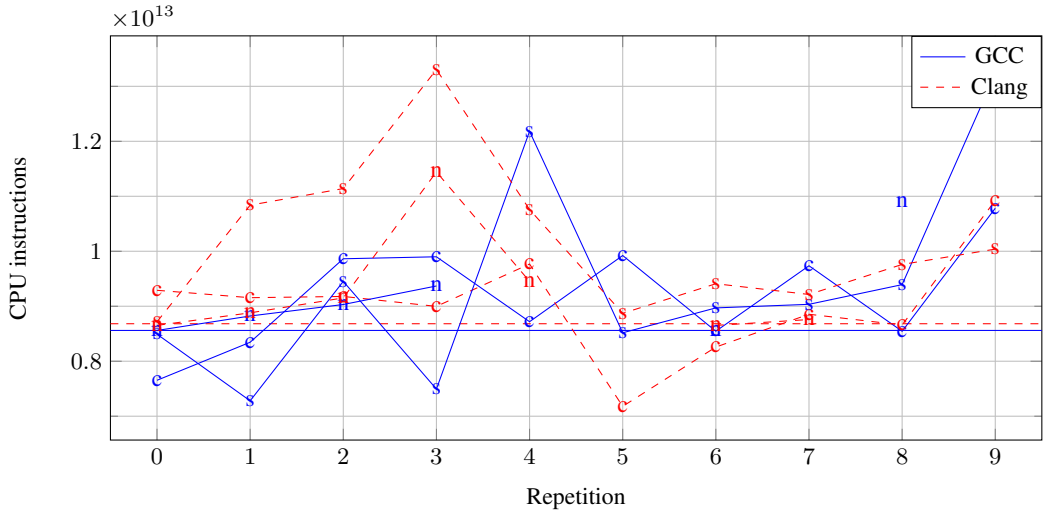
\begin{figure}[tb!]
  \small
  \centering
  \begin{tikzpicture}
    \begin{axis}[
        scale only axis,
        ymin=0,
        ymax=1.75e13,
        xmin=-0.5,
        xmax=9.5,
        minor y tick num=1,
        tick scale binop=\times,
        grid=both,
        xtick={0,1,2,3,4,5,6,7,8,9},
        xlabel={Repetition},
        ylabel={CPU instructions},
        width=0.75\linewidth,
        height=17em,
      	legend style={at={(1,1)},anchor=north east}]
      \addplot[color=blue,mark=none] coordinates {
        (-0.5, 13884363297634)
        (9.5, 13884363297634)
      };
      \addplot[color=red,dashed,mark=none] coordinates {
        (-0.5, 12678457819901)
        (9.5, 12678457819901)
      };
      \addplot[color=blue,dashed,mark=text,text mark=c] coordinates {
        (0, 1052233259935)
        (1, 748605970561)
        (2, 683986025142)
        (3, 714253077313)

        (5, 497886536003)
        (6, 936227516526)
        (7, 817146779869)

        (9, 508194555575)
      };
      \addplot[color=red,dashed,mark=text,text mark=c] coordinates {
        (0, 624722390794)
        (1, 497172398619)
        (2, 644991705932)
        (3, 662309325965)

        (5, 571548157996)
        (6, 940629814218)
        (7, 777828940088)

        (9, 423730660870)
      };
      \addplot[color=blue,dashed,mark=text,text mark=s] coordinates {
        (1, 9020575227822)
        (2, 5354668585551)

        (8, 5647635996589)
      };
      \addplot[color=red,dashed,mark=text,text mark=s] coordinates {
        (1, 8322424758490)
        (2, 12675459569063)
        (3, 12586021712076)
        (4, 11455506713413)
        (5, 11201088782847)
      };
      \addplot[color=blue,dashed,mark=text,text mark=n] coordinates {
        (0, 13783278364202)

        (3, 13777559869118)

        (5, 8902963497574)

        (9, 13637231623338)
      };
      \addplot[color=red,dashed,mark=text,text mark=n] coordinates {
        (0, 12639860591441)

        (3, 11813590294820)

        (5, 8226681357754)
      };
      \legend{GCC,Clang}
    \end{axis}
  \end{tikzpicture}
  \caption{Algorithm configuration and source code evolution using MAGPIE on the LPG scenario. (c:~configuration, s:~statements, n:~numerical constants)}
  \label{tikz:test_lpg}
\end{figure}
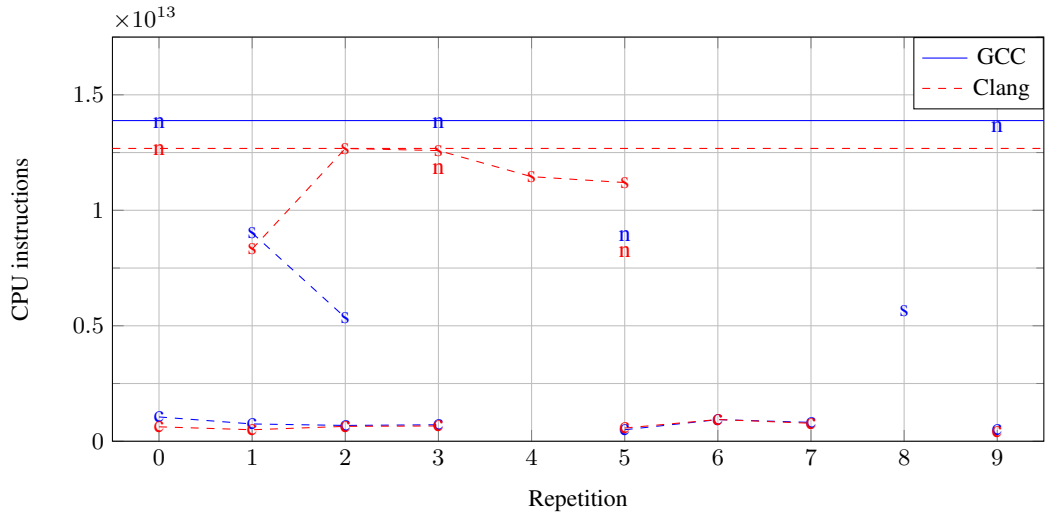
\begin{figure}[tb!]
  \small
  \centering
  \begin{tikzpicture}
    \begin{axis}[
        scale only axis,
        xmin=-0.5,
        xmax=9.5,
        ytick distance=0.5e13,
        minor y tick num=1,
        tick scale binop=\times,
        grid=both,
        xtick={0,1,2,3,4,5,6,7,8,9},
        xlabel={Repetition},
        ylabel={CPU instructions},
        width=0.75\linewidth,
        height=17em,
      	legend style={at={(1,1)},anchor=north east}]
      \addplot[color=blue,mark=none] coordinates {
        (-0.5, 41658295567448)
        (9.5, 41658295567448)
      };
      \addplot[color=red,dashed,mark=none] coordinates {
        (-0.5, 44268098376402)
        (9.5, 44268098376402)
      };
      \addplot[color=blue,mark=text,text mark=c] coordinates {
        (0, 41609583857100)
        (1, 51869490084252)
        (2, 41482955831232)
        (3, 40595539922549)
        (4, 41563695595816)
        (5, 41790786935462)
        (6, 50141598868937)
        (7, 42197543162309)
        (8, 41532349842586)
        (9, 35931741771636)
      };
      \addplot[color=red,dashed,mark=text,text mark=c] coordinates {
        (0, 47213862751244)
        (1, 42613639368699)
        (2, 43748817503162)
        (3, 44674369224988)
        (4, 46607078630641)
        (5, 43819335873878)
        (6, 54797638468384)
        (7, 44499498759452)
        (8, 45944657306352)
        (9, 44215614377174)
      };
      \addplot[color=blue,mark=text,text mark=s] coordinates {
        (0, 41589611007616)
        (1, 41903568354490)
        (2, 40929321108396)
        (3, 40637008609400)
        (4, 41300194603274)
        (5, 41777904893932)
        (6, 41518676320837)
        (7, 41454714028336)
        (8, 41756183385902)
        (9, 41607135872459)
      };
      \addplot[color=red,dashed,mark=text,text mark=s] coordinates {
        (0, 46367310733109)
        (1, 44143663176156)
        (2, 44094447260999)
        (3, 46274335156428)
        (4, 43840854679871)
        (5, 43622625846549)
        (6, 46336249617821)
        (7, 43748098782335)
        (8, 44174603725790)
        (9, 44489380548375)
      };
      \addplot[color=blue,mark=text,text mark=n] coordinates {
        (0, 41918360709241)
        (1, 41626979846607)
        (2, 41430193918584)
        (3, 41729443005330)
        (4, 41662334692251)
        (5, 41680630856436)
        (6, 41484111151360)
        (7, 41625786789976)
        (8, 41779758667642)
        (9, 41664761031319)
      };
      \addplot[color=red,dashed,mark=text,text mark=n] coordinates {
        (0, 46300321347445)
        (1, 44110214629702)
        (2, 43981326694348)
        (3, 46330452828743)
        (4, 45560397542114)
        (5, 43271402612880)
        (6, 45348354474294)
        (7, 44016733131589)
        (8, 44103938684954)
        (9, 43984993552031)
      };
      \legend{OpenJDK,GraalVM}
    \end{axis}
  \end{tikzpicture}
  \caption{Algorithm configuration and source code evolution using MAGPIE on the Sat4J scenario. (c:~configuration, s:~statements, n:~numerical constants)}
  \label{tikz:test_sat4j}
\end{figure}
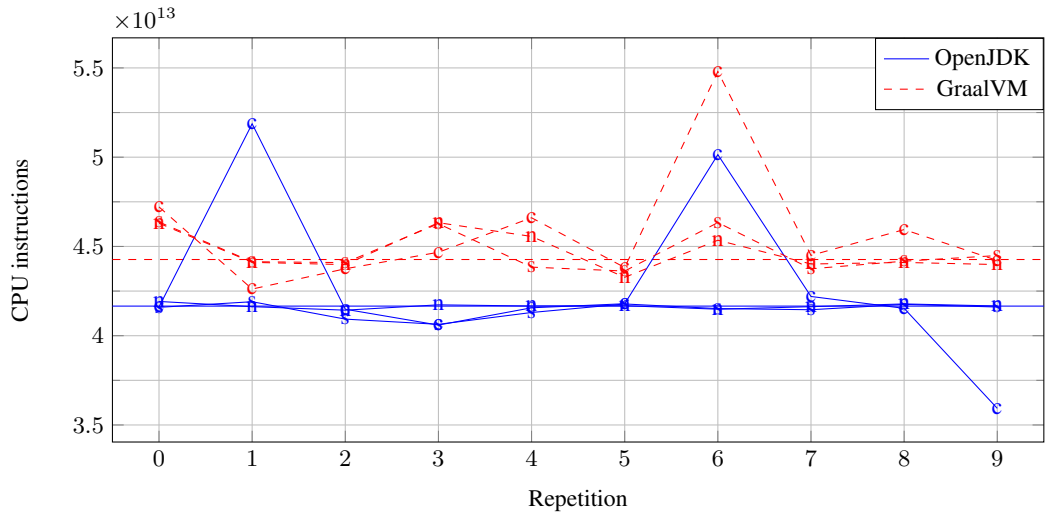
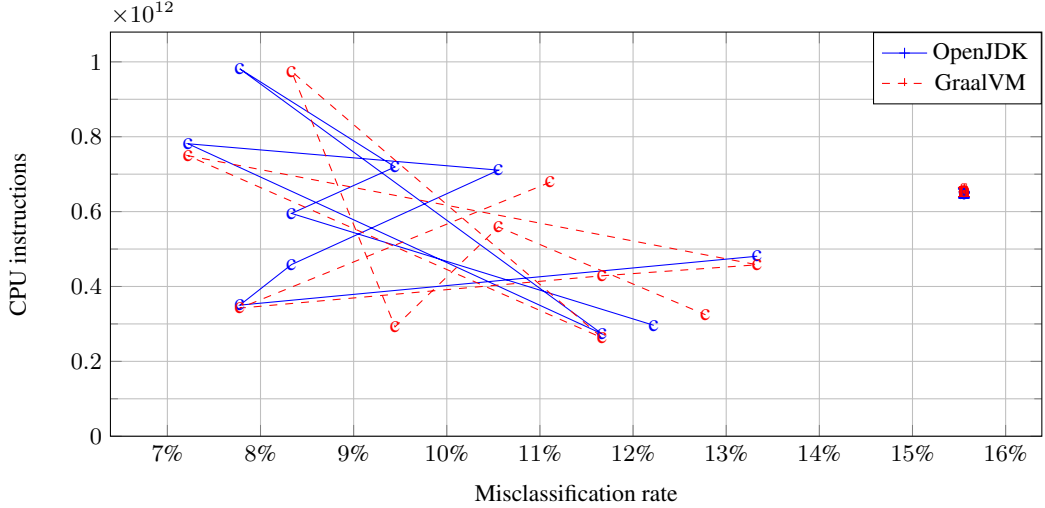
\begin{figure}[tb!]
  \small
  \centering
  \begin{tikzpicture}
    \begin{axis}[
        scale only axis,
        x filter/.code={\pgfmathparse{#1/180}\pgfmathresult},
        xticklabel={\pgfmathparse{\tick*100}\pgfmathprintnumber{\pgfmathresult}\%},
        ymin=0,
        minor y tick num=1,
        tick scale binop=\times,
        grid=both,
        xlabel={Misclassification rate},
        ylabel={CPU instructions},
        width=0.75\linewidth,
        height=17em,
      	legend style={at={(1,1)},anchor=north east}]
      \addplot[color=blue,mark=+] coordinates {
        (28, 654365490511)
        (28, 651992914194)
        (28, 652497552193)
        (28, 650971515130)
        (28, 649681595357)
        (28, 651787715416)
        (28, 648318168964)
        (28, 652431271734)
        (28, 650119815179)
        (28, 652082961223)
      };
      \addplot[color=red,dashed,mark=+] coordinates {
        (28, 659239693600)
        (28, 652726688739)
        (28, 653081861454)
        (28, 654424550102)
        (28, 659987364418)
        (28, 653529524439)
        (28, 653882842981)
        (28, 653261883418)
        (28, 652621740915)
        (28, 653766229520)
      };
      \addplot[color=blue,mark=text,text mark=c] coordinates {
        (24, 480855352411)
        (14, 350171807276)
        (15, 458616309621)
        (19, 710751026594)
        (13, 781220321649)
        (21, 273517729034)
        (14, 981387606711)
        (17, 719297377888)
        (15, 595153657806)
        (22, 296288141409)
      };
      \addplot[color=red,dashed,mark=text,text mark=c] coordinates {
        (20, 678581190595)
        (14, 342210657663)
        (21, 428547678429)
        (24, 457946326881)
        (13, 749305305567)
        (21, 263025912880)
        (15, 974601985183)
        (17, 292136554643)
        (19, 559087233668)
        (23, 324526747291)
      };
      \addplot[color=blue,mark=text,text mark=s] coordinates {
        (28, 649861481135)
        (28, 650093844326)
        (28, 649481384212)
        (28, 648325208268)
        (28, 649811955324)
        (28, 648213220413)
        (28, 650639873644)
        (28, 648146672186)
        (28, 649240478058)
        (28, 649938104989)
      };
      \addplot[color=red,dashed,mark=text,text mark=s] coordinates {
        (28, 652881741840)
        (28, 660409303911)
        (28, 653403513275)
        (28, 654079087120)
        (28, 654723131836)
        (28, 653475010037)
        (28, 653207865579)
        (28, 653195292697)
        (28, 653224104595)
        (28, 653685297675)
      };
      \addplot[color=blue,mark=text,text mark=n] coordinates {
        (28, 649690052124)
        (28, 648809915877)
        (28, 650003466497)
        (28, 650477946916)
        (28, 648553498274)
        (28, 649834724492)
        (28, 650413498012)
        (28, 648154147157)
        (28, 649679153615)
        (28, 649890506348)
      };
      \addplot[color=red,dashed,mark=text,text mark=n] coordinates {
        (28, 653776468120)
        (28, 654018091964)
        (28, 653555288927)
        (28, 653283192273)
        (28, 652839633193)
        (28, 653149930602)
        (28, 653373543190)
        (28, 653097208202)
        (28, 652649274157)
        (28, 654451724083)
      };
      \legend{OpenJDK,GraalVM}
    \end{axis}
  \end{tikzpicture}
  \caption{Algorithm configuration and source code evolution using MAGPIE on the Weka scenario. (c:~configuration, s:~statements, n:~numerical constants)}
  \label{tikz:test_weka}
\end{figure}

Figures~\ref{tikz:test_minisat}, \ref{tikz:test_lpg}, \ref{tikz:test_sat4j}, and \ref{tikz:test_weka} report the fitness values on the set of test instances (respectively for the MiniSAT, LPG, Sat4j, and Weka scenarios) of the ten evolved mutants obtained with MAGPIE using algorithm configuration (denoted by~``c''), GI on statements~(``s'') and GI on numeric literals~(``n'').
Lines connect repeated runs to emphasise differences obtained between alternative compilers/JVM.
Missing data points correspond to overfitting mutants --- i.e., mutants for which the validation step failed to produce a single mutant improving the baseline.
Data points above the baseline correspond to mutants both better during training and validation, but failing to properly generalise on the test set.
Note that results may be biased toward algorithm configuration as scenarios and datasets have been lifted from AClib, a configuration library.
Differences with other work may also be due to the specific configuration encoding, the simple local search used, or training protocol.

In the MiniSAT scenario (\autoref{tikz:test_minisat}) most mutants fail to generalise and very few mutants achieve improvements over the baseline fitness (using \texttt{-O3}), noticeably two by changing MiniSAT configuration and two by evolving its AST statements.
In the LPG scenario (\autoref{tikz:test_lpg}) while all data points show improvements over the baseline, the majority of GI-evolved mutants overfit during the validation step.
Even if both GI types of approaches resulted in some very good mutants, with up to 60\% speedup, the best improvements achieved are obtained using algorithm configuration, with a best fitness of 3.42\% over the associated baseline fitness.
In the Sat4j scenario (\autoref{tikz:test_sat4j}) all three types of approaches similarly show little improvement over the default setup, with two tuned mutants and one statement-evolved mutant failing to generalise.
Finally \autoref{tikz:test_weka} reports results for the Weka scenario, showing both the number of CPU instructions and the misclassification rate over the test dataset (both to be minimised).
Neither the evolution of statement or numerical values within the random forest implementation led to any substantial improvement (all mutants are clustered with the baseline with a misclassification rate of 15.5\%).
On the other hand, tuning of the algorithm's parameters led to improvements in both objectives, with in particular variants being simultaneously twice as fast and twice as efficient as the baseline.

\begin{table*}[t!]
  \caption{Single best fitness improvement on test instances. [in percentage of CPU instructions]}
  \centering
  \begin{tabular}{l ccccccc}
    \toprule
    Scenario & Compiler & Parameters & Statements & Numbers & Combined & Joint\\
    \midrule
    MiniSAT (GCC) & -9\% & -11\% & -15\% & -0\% & -25\% & -40\%\\
    MiniSAT (Clang) & -3\% & -17\% & -0\% & -0\% & -21\% & -10\%\\
    LPG (GCC) & -12\% & -97\% & -61\% & -36\% & -95\% & -94\%\\ 
    LPG (Clang) & -0\% & -97\% & -58\% & -35\% & -95\% & -0\%\\ 
    Sat4j (Oracle) & -19\% & -13\% & -1\% & -0\% & -29\% & -8\%\\
    Sat4j (GraalVM) & -25\% & -4\% & -1\% & -1\% & -27\% & -24\%\\
    Weka (Oracle) & (-0\%, -5\%) & (-50\%, -46\%) & (-0\%, -0\%) & (-0\%, -0\%) & (-54\%, +15\%)$^1$ & (-50\%, +13\%)$^2$\\
    Weka (GraalVM) & (-0\%, -4\%) & (-54\%, +14\%)$^3$ & (-0\%, -0\%) & (-0\%, -0\%) & (-54\%, +10\%)$^4$ & (-57\%, -36\%)\\
    \bottomrule
    \label{table:combined}
  \end{tabular}\\
  For Weka are reported changes for both misclassification rate and CPU instructions.\\
  Other notable Weka improvements: $^1$: (-50\%, -49\%), $^2$: (-43\%, -24\%), $^3$: (-50\%, -48\%), $^4$: (-50\%, -49\%)
\end{table*}

\autoref{table:combined} reports, within the ten runs of each approach, the single best performance improvement.
For Weka, for which were to be minimised (in lexicographical order) both misclassification rate and number of CPU instructions, fitness values of some additional runs are also reported when relevant.
The results relative to the ``combined'' and ``joint'' columns are discussed in the following sections.

Algorithm configuration is the clear winner in the LPG and Weka scenarios.
For LPG, it achieves up to 97\% improvement in number of CPU instructions, i.e., the software variant is around 30 times faster, when GI \emph{only} produces up to 61\% improvement.
For Weka, GI approaches are unable to find any improvements.
One reason may be its very concise programming style, with very small dedicated functions hard to safely modify.
Another reason could be that improvements found by algorithm configuration may fall outside the scope of the targeted classifier source code, within the more general code not considered by GI.
For MiniSAT  all three improvement techniques resulted in significant speedups (9\%, 17\%, and 15\%, respectively).
Finally, for Sat4j, both compiler optimisation and algorithm configuration were able to find significant speedups, (25\% and 13\%, respectively).

There is, however, no single technique best in every scenario.
By nature, algorithm configuration explores design choices specifically exposed by software designers, but are limited to those exclusively.
On the other hand, genetic improvement produces changes from a much larger and much richer search space, well outside what can reasonably be manually described with parameters, but can easily suffer from focusing on an inadequate part of the fitness landscape.

\smallskip
\centerline{
\noindent
\fbox{
  \parbox{0.96\linewidth}{
    \textbf{Answer to RQ2 and RQ3:}
    MAGPIE was successful in finding large running time improvements, in each of the three search spaces investigated: on the LPG scenario up to 97\% speedup for parameter optimisation, up to 61\% speedup for GI on statements, and up to 36\% speedup for GI on numerical constants.
  }
}
}

\subsection{RQ4: Evaluating Combinations of Mutants}
\label{subsection:rq3}

In~\autoref{table:combined} we report the performance on the validation set of software variants obtained by combining edits found by MAGPIE during training of all four types of approaches.
Changes in compiler or interpreter configuration are unquestionably not reachable through the other search spaces, as they don't impact the software source code or execution semantics directly.
As such, it seems that associated improvements directly lead to matching improvements in the combined patch.
This is particularly visible in the MiniSAT (Clang), Sat4j, and Weka scenarios, in which there is no interference between algorithm configuration and genetic improvement.

In the LPG scenario, the combined validation is slightly less efficient than the pure configuration software variant (from 97\% to 95\% speedup).
In the Weka scenario, it successfully manages to combine the 5\% configuration changes from the JVM optimisation to the improvements found using Weka's algorithm configuration, resulting in slightly better variants.
In both MiniSAT and Sat4j scenarios, however, it successfully combines the improvement found in the different search spaces and leads to significantly better software variants, ranging from further 2\% to 10\% speedups.

\begin{table}[t!]
  \caption{Percentage of edits by origin in the best combined and joint patches from \autoref{table:combined} (detailing between {\upshape\texttt{StmtDelete}/\allowbreak\texttt{StmtReplace}/\allowbreak\texttt{StmtInsert}} and {\upshape\texttt{ConstantSet}/\allowbreak\texttt{ConstantUpdate}})}
  \centering
  \begin{tabular}{l ccc@{\,+}c@{\,+\!}cc@{\,+\!\!}c}
    \toprule
    Scenario & Comp. & Param. & \multicolumn{3}{c}{Statements} & \multicolumn{2}{c}{Numbers}\\
    \midrule
    \multicolumn{3}{l}{\textbf{Combined validation}}\\
    MiniSAT (GCC) & 17 & 9 & \bf 34 & 9 & 0 & 16 & 5\\
    MiniSAT (Clang) & 6 & 14 & \bf 43 & 6 & 6 & 9 & 17\\
    LPG (GCC) & 17 & \bf 25 & \bf 25 & 10 & 4 & 7 & 13\\ 
    LPG (Clang) & 0 & 34 & \bf 39 & 2 & 5 & 10 & 10\\ 
    Sat4j (Oracle) & \bf 48 & 10 & 7 & 3 & 3 & 0 & 8\\
    Sat4j (GraalVM) & \bf 47 & 18 & 0 & 12 & 6 & 6 & 12\\
    Weka (Oracle) & \bf 44 & 9 & 9 & 3 & 9 & 13 & 13\\
    Weka (GraalVM) & \bf 44 & 12 & 12 & 4 & 16 & 8 & 4\\
    \cmidrule(lr){1-8}
    \multicolumn{3}{l}{\textbf{Joint training}}\\
    MiniSAT (GCC) & 15 & 8 & \bf 42 & 8 & 5 & 14 & 9\\
    MiniSAT (Clang) & 3 & 11 & \bf 40 & 20 & 0 & 17 & 9\\
    LPG (GCC) & 0 & 14 & \bf 40 & 0 & 11 & 11 & 11\\ 
    LPG (Clang) & 0 & 0 & 0 & 0 & 0 & 0 & 0\\ 
    Sat4j (Oracle) & 0 & \bf 33 & 0 & \bf 33 & 0 & 0 & \bf 33\\
    Sat4j (GraalVM) & \bf 29 & 0 & 14 & 14 & 0 & 14 & \bf 29\\
    Weka (Oracle) & 0 & 20 & 7 & \bf 27 & 7 & 20 & 20\\
    Weka (GraalVM) & 29 & \bf 43 & 0 & 14 & 0 & 14 & 0\\
    \bottomrule
    \label{table:composition}
  \end{tabular}
\end{table}

\autoref{table:composition} presents the composition of the combined (RQ3) and joint (RQ4) patches reported in \autoref{table:combined}.
In the MiniSAT and LPG scenarios best improvements come from statement deletion. 
In the Sat4j and Weka scenarios best improvements are found through compiler/interpreter configuration, while combined patches make use of every type of edits.

\smallskip
\noindent
\centerline{
\fbox{
  \parbox{0.96\linewidth}{
    \textbf{Answer to RQ4:}
    Edits resulting from different techniques can be effectively combined, leading up to further 10\% speedups when compared to the best individual speedup.
    Detailed analysis of combined patches composition hints that optimal software variants may only be reached using a diverse set of edits.
  }
}
}

\subsection{RQ5: Simultaneous Evolution}
\label{subsection:rq4}

Performance of software variants obtained by simultaneously training on all four search spaces --- i.e., optimising compiler/JVM configuration, algorithm configuration, GI on statements and GI on numbers --- is also reported in \autoref{table:combined}, together with composition details in \autoref{table:composition}.

Despite having a similar training budget as the combined approach discussed in the previous section, joint training produced worse single best results in five of the eight scenarios.
However, in two cases, optimising MiniSAT using GCC and optimising Weka using GraalVM, the experiment produced a significantly better software variant.
For MiniSAT it led to a 40\% speedup when the best combined and best individual patches only produced 25\% and 15\% speedups, respectively.
For Weka joint training produced a variant with both a 57\% decrease in misclassification rate (the best achieved in our results) with a 36\% speedup.
This can be explained as the joint search space is more complex to navigate, while providing opportunities to find otherwise inaccessible combinations of changes.
Finally, we note that final patches obtained using joint training seem to use a more diverse set of edits, with more edits associated with each of the three improvement techniques.

The results show that the joint training clearly suffers from the size of the combined search space.
To alleviate this, a possible solution may be to change how search spaces are accessible throughout the search.
For example, the search could first strongly focus on compiler or interpreter configuration, as those changes seem to be mostly independent from the others, before introducing other edit types.
The relative probabilities to investigate specific types of changes may also be specifically tuned~\cite{smigielska:2021:gi-icse} to favour some search spaces over others; follow a set schedule~\cite{pageau:2019:emo}; or use a multi-armed bandits scheme to maximise expected improvements.

\smallskip
\noindent
\centerline{
\fbox{
  \parbox{0.96\linewidth}{
    \textbf{Answer to RQ5:}
    Training on the joint search space proved possible and effective.
    It resulted in a MiniSAT software variant 25\% faster than the best variant found by any technique applied separately, and still 15\% faster than the variant obtained by combining the best final patches.
    However, it was less efficient in the other scenarios.
  }
}
}

\section{Conclusions}
\label{section:conclusion}

In this paper we proposed a unified software performance improvement framework: MAGPIE.
It provides a shared solution representation, a sequence of edits.
We implemented three software performance improvement techniques using the MAGPIE approach: compiler optimisation, algorithm configuration, and genetic improvement.
We evaluate our framework on four large real-world software, targeting running time minimisation and a bi-objective scenario involving running time and solution quality.
Results of our experiments are threefold.
First, we showed that even a simple local search can find significant performance improvement for all three techniques: up to 25\% for compiler optimisation, 97\% for algorithm configuration and 61\% for evolving source code using genetic improvement.
Then, we discussed the best variants found, and show that while no single technique is most efficient in all scenarios, partial recombination of individually evolved changes led to up to 10\% further speedups.
Finally, we investigated a novel approach in which all three techniques are combined during the local search itself, and were able to find an additional 15\% speedup in one specific scenario.

Overall, we implemented a simple alternative to existing dedicated tooling to automate software performance improvement.
Our results show that our framework is both effective and efficient in reproducing significant improvements in line with previous work using other representations and search approaches.
It provides a practical interface to conduct comparison studies of both improvement techniques and their associated search processes, and allows for efficient prototyping of new research.

In future work we intend to investigate more closely the fitness landscapes associated within the different search spaces.
We also plan to extend our framework with new and more complex search processes, new software improvement techniques, as well as support for other performance measures.
Finally, we intend to simplify and streamline our framework to provide the community with both a comprehensive and extendable toolkit for researchers and a readily usable framework for practitioners.

\textbf{Implementation:} \url{https://github.com/bloa/magpie}

\textbf{Acknowledgements:} This work was supported by UK EPSRC Fellowship EP/P023991/1.

\bibliographystyle{ACM-Reference-Format}
\bibliography{bib_longest}

\end{document}